# The Stablecoin Discount: Evidence of Tether's U.S. Treasury Bill Market Share in Lowering Yields

**Lennart Ante**, a, b, *, **Aman Saggu** c, b,, **Ingo Fiedler** b, d

*a Constructor University, School of Business, Social & Decision Sciences, Germany*
*b Blockchain Research Lab, Germany*
*c Mahidol University, Business Administration Division, Mahidol University International College*
*d Concordia University, Canada*

\* *Correspondence: lante@constructor.university*

*This version*: May 18th, 2025

**Abstract:** Stablecoins represent a critical bridge between cryptocurrency and traditional finance, with Tether (USDT) dominating the sector as the largest stablecoin by market capitalization. By Q1 2025, Tether directly held approximately $98.5 billion in U.S. Treasury bills, representing 1.6% of all outstanding Treasury bills, making it one of the largest non-sovereign buyers in this crucial asset class, on par with nation-state-level investors. This paper investigates how Tether's market share of U.S. Treasury bills influences corresponding yields. The baseline semi-log time trend model finds that a 1% increase in Tether's market share is associated with a 1-month yield reduction of 3.8%, corresponding to 14-16 basis points. However, threshold regression analysis reveals a critical market share threshold of 0.973%, above which the yield impact intensifies significantly. In this high regime, a 1% market share increase reduces 1-month yields by 6.3%. At the end of Q1 2025, Tether's market share placed it firmly within this high-impact regime, reducing 1-month yields by around 24 basis points relative to a counterfactual. In absolute terms, Tether's demand for Treasury Bills equates to roughly $15 billion in annual interest savings for the U.S. government. Aligning with theories of liquidity saturation and nonlinear price impact, these results highlight that stablecoin demand can reduce sovereign funding costs and provide a potential buffer against market shocks.

# 1  Introduction

The emergence of stablecoins represents a transformative shift at the intersection of cryptocurrency and traditional finance, evolving from a niche innovation (Ante et al., 2021; Griffin and Shams, 2020; Saggu, 2022) to widely used financial instruments (Ante, 2025). By May 2025, stablecoin market capitalization exceeded USD 240 billion, reflecting rapid adoption as mediums of exchange, stores of value, and critical liquidity anchors in the decentralized finance (DeFi) ecosystem (CoinGecko, 2025). Stablecoins maintain a stable value relative to a reference asset using full or partial collateralization, algorithmic controls, or hybrid models, drawing value from reserves or other stabilization mechanisms (Maex and Slavov, 2025). Tether (USDT) dominates the stablecoin market with over USD 150 billion in circulation (~62.5% stablecoin market share), maintained at a 1:1 peg to the U.S. dollar and backed by an overcollateralized pool of assets—primarily USD 98.5 billion in U.S. Treasury bills (BDO, 2025), representing around 1.6% of all outstanding $6.16 trillion U.S. Treasury bills (Department of the Treasury, 2025). The scale of market participation in the world's most liquid and systemically important financial market positioned Tether among the largest non-sovereign buyers of U.S. Treasuries in 2024, comparable to nation-state-level investors (Sandor, 2025). This raises critical questions about the impact of stablecoin demand for U.S. Treasuries on corresponding sovereign debt yields and financing costs of government debt (Azar et al., 2024).

A substantial body of literature has established that large institutional investors can influence Treasury yields through their trading behavior. Preferred habitat theory, as articulated by Greenwood and Vayanos (2014), explains that investors with strong maturity preferences can influence specific yield curve segments and thereby the term structure of interest rates. Similarly, Lou et al. (2013) demonstrate that price effects intensify when demand clusters around Treasury auction cycles, consistent with the theory of segmented markets. Krishnamurthy and Vissing-Jorgensen (2012) further provide evidence that demand shifts for safe assets can materially affect Treasury yields by altering the relative scarcity of government securities, consistent with theories emphasizing the role of fixed income supply-demand imbalances. Nagel (2016) shows that regulatory shocks can reshape yield curves by altering the liquidity premium on near-money assets, consistent with theories highlighting the role of market frictions and liquidity constraints in fixed income pricing. However, the influence of stablecoin issuers like Tether—a rapidly growing class of market participants—remains largely unexplored, despite their significant presence as major non-sovereign buyers in the U.S. Treasury market.

Against this backdrop, we propose two hypotheses: (H1): As Tether's market share of outstanding U.S. Treasury bills increases, it exerts downward pressure on short-term yields. This hypothesis is motivated by price pressure effects observed for large institutional investors, where concentrated demand increases the relative scarcity of safe assets, compresses yields, and intensifies liquidity constraints (Greenwood and Vayanos, 2014; Krishnamurthy and Vissing-Jorgensen, 2012; Lou et al., 2013; Nagel, 2016). (H2): Tether's acquisitions of U.S. Treasury bills induce disproportionately larger yield reductions once its holdings surpass a critical market share threshold. This hypothesis is motivated by evidence that large, concentrated buyers induce nonlinear price effects, where sustained demand pressure in segmented markets drives disproportionately larger price responses as supply constraints become binding (Gabaix et al., 2006; Kyle, 1985).

To test the first hypothesis (H1), we estimate a semi-log model to analyze the impact of Tether's market share of 1-month and 3-month outstanding U.S. Treasury bills on corresponding yields. Firstly, we estimate a baseline model without controls, revealing a weak positive relationship, indicative of broader macroeconomic and monetary factors obscuring the relationship. Secondly, we introduce a time trend to capture these longer-term dynamics, reversing the relationship and yielding statistically significant semi-elasticities of 4.144 (-3.376) for 1-month (3-month) yields. Thirdly, we estimate fully specified models further accounting for residualized U.S. Treasury bill issuance, yielding semi-elasticities of 3.795 (-3.386) for 1-month (3-month) yields, demonstrating that a 1% increase in Tether's market share reduces yields by approximately 3.8% (Missing), or 14-16 (Missing) basis points respectively, reflecting substantial direct liquidity effects, consistent with H1.

To test the second hypothesis (H2), we apply a threshold regression model to identify the point at which the impact of Tether's market share on yields shifts. Firstly, we conduct a grid search to determine the optimal threshold, identifying 0.973% as the critical level where Tether's influence on yields changes significantly. Secondly, we estimate the model and find that in the low regime, where Tether's market share is 0.973% or less, a 1% increase in market share reduces 1-month (3-month) yields by 1.73% (1.42%), indicating modest but statistically significant downward pressure. Lastly, in the high regime, where Tether's market share exceeds this threshold, a 1% increase reduces 1-month (3-month) yields by 6.26% (5.82%), reflecting disproportionately larger yield impacts as Tether's share grows, aligning with theoretical models of liquidity and price impact, where concentrated buyers exert escalating downward pressure on yields as their market share surpasses critical levels, consistent with H2.

## 2  Data and Descriptive Statistics

The dataset summarized in Table 1 spans Q1 2022 to Q1 2025, a period characterized by rapid global stablecoin growth (Feingold, 2025). It includes Tether's U.S. Treasury bill holdings, sourced from independently verified public disclosures and attestation reports, which provide precise reserve composition data.[1] [2] We compute (iii) Tether's market share as a monthly percentage of total outstanding U.S. Treasury bills, utilizing U.S. Department of the Treasury data, which regularly reports the composition and maturity of government debt. Tether's market share increased from 0.91% in Q1 2022 to 1.60% in Q1 2025, reflecting its rising influence in short-term sovereign debt markets and its broader role as a liquidity anchor in both digital asset and traditional financial markets, aligning with wider trends in stablecoin adoption. Tether's market share over the sample period averages 1.07%, ranging from 0.57% to 1.60%.

We analyze 1-month and 3-month U.S. Treasury yields as the primary outcome variables, consistent with Tether's disclosure that its U.S. Treasury bill holdings have a maximum maturity of 90 days. These yields, drawn from the Federal Reserve Economic Data (FRED) database, capture secondary market rates for newly issued 4-week and 13-week U.S. Treasury bills, corresponding to 1-month and 3-month maturities, respectively. Given the high sensitivity of U.S. Treasury yields to monetary policy shifts, reflecting shifts in investor expectations and liquidity conditions, we log-transform these yields to (i) Log 1-Month Yields and (ii) Log 3-Month Yields. This approach, standard in fixed-income analysis, improves interpretability, reduces skewness, stabilizes variance, and supports elasticity-based interpretations, allowing for more precise estimation of relative yield changes in response to market forces (Hanson and Stein, 2015).[3] The (i) Log 1-month yields—a key proxy for short-term funding costs—averaged 4.16%, fluctuating from a low of 0.52% in early 2022 to a peak of 5.54% in early 2025.

The (v) T-Bill Changes IHS variable measures the monthly change in total outstanding U.S. Treasury bills, sourced from U.S. Department of the Treasury data. To address large values and accommodate both positive and negative changes, we utilize an inverse hyperbolic sine (IHS) transformation defined as: $IHS(x) = \ln(x + \sqrt{x^2 + 1}$. This approach retains the interpretability benefits of a log-like scale while reducing heteroskedasticity and mitigating the influence of potential outliers (Bellemare and Wichman, 2020). To inform the model specification, we examine the correlation structure among variables. The correlation matrix reveals that (iii) Tether's market share positively correlates with both the (i) Log 1-month (0.48) and (ii) Log 3-month (0.46) yields, which is initially counterintuitive as higher demand typically lowers yields. However, this positive relationship reflects broader monetary factors than a direct demand effect. (iii) Tether's market share is also strongly correlated with (v) T-Bill Changes IHS (0.94) and moderately with the time trend (0.64), capturing the Federal Reserve's aggressive rate-hiking cycle during the sample period. To ensure the robustness of our analysis, all subsequent model specifications include the (iv) Time Trend variable. This inclusion is crucial as it controls for underlying monetary conditions, reassuring that the estimated relationship between Tether's holdings and U.S. Treasury bill yields captures more direct effects rather than spurious associations driven by broader monetary shifts.

---

[1] Tether's Consolidated Reserves Reports (CRRs), available on the Tether Transparency page (https://tether.to/en/transparency/?tab=reports), began with consulting memoranda from Friedman LLP on September 28, 2017, without AICPA assurance. In June 2018, Freeh, Sporkin & Sullivan LLP (FSS) provided a legal review confirming bank balances. From March 2021 to December 2023, Moore Cayman conducted quarterly assurance reports under ISAE 3000 (Revised). Since March 2024, BDO Italia S.p.A. has provided assurance under the same standard. Tether first disclosed U.S. Treasury bill holdings separately on September 30, 2021. The June 30, 2022, report fully separated U.S. Treasury bills from commercial paper. By September 30, 2023, Tether included direct U.S. Treasury bill holdings and indirect exposure through money market funds and reverse repurchase agreements. The March 31, 2025, report provided the most detailed breakdown, listing over $94 billion in U.S. Treasury bills with average maturities under 90 days.

[2] All models and results in this study rely exclusively on Tether's direct U.S. Treasury bill holdings, excluding indirect exposures through money market funds and similar vehicles. As of March 31, 2025, Tether held approximately $20 billion U.S. Treasuries indirectly, $4.88 billion through money market funds, and $15.09 billion through overnight reverse purchase agreements. Our analysis also focuses on 1-month and 3-month U.S. Treasury bills, consistent with Tether's disclosures that its Treasury holdings have a residual average maturity of less than 90 days. Including these indirect holdings would imply even larger reductions in U.S. government borrowing costs (BDO, 2025).

[3] During the sample period, the Federal Reserve implemented aggressive monetary tightening to combat inflation, raising rates from 0.25%–0.50% in March 2022 (Q1 2022) to a peak of 5.25%–5.50% in July 2023 (Q3 2023), before gradually reducing them to 4.25%–4.50% from September 2024 to April 2025 (Q3 2024 to Q1 2025). These changes in the Federal Reserve's policy had a direct impact on U.S. Treasury bill yields.

To address potential multicollinearity, we generate (vi) T-Bill Changes Residual by regressing (v) T-Bill Changes IHS on (iv) Time Trend and (iii) Tether's market share, then scaling the residuals by 1,000 for readability. This process isolates the unique component of outstanding U.S. Treasury bill changes not linearly explained by the (iv) Time Trend or (iii) Tether's market share. This residualization step reduces variance inflation, mitigates the risk of biased coefficient estimates, and improves interpretability by ensuring that this control variable captures only the unexplained portion of outstanding Treasury bill changes, enhancing the precision and robustness of our main regression models.

## 3 Results

### 3.1 Baseline Time Trend Model

Table 2 presents the results of the baseline time trend model, defined in Equation (1), which estimates the relationship between Tether's market share and U.S. Treasury bill yields, employing a semi-log specification to capture the elasticity of yields with respect to market share, accounting for both long-term monetary trends and short-term issuance shocks.

$$\ln(y_t) = \alpha + \beta(S_t) + \gamma T_t + \delta R_t + \varepsilon_t \qquad (1)$$

where $y_t$ is the log of (i) 1-month yields in models (1) to (3) or log of (ii) 3-month yield in models (4 to 6), $S_t$ defines (iii) Tether's market share, $T_t$ denotes the linear (iv) time trend, $R_t$ represents the (vi) T-bill changes residual variable, and $\varepsilon_t$ is the error term. This specification isolates the direct impact of (iii) Tether's market share by controlling for broader monetary trends and short-term supply shocks. For each yield maturity, we estimate three models: (1 and 4) a baseline model without controls, (2 and 5) a model including the (iv) time trend to account for monetary dynamics, and (3 and 6) a fully specified model incorporating both the (iv) time trend and the (vi) residualized T-bill issuance variable to account for short-term supply shocks. The coefficient $\beta(S_t)$ measures the semi-elasticity of a percentage change in yield for a one-unit change in Tether's market share, critical to understanding the "crowding out" effect of Tether's rapid accumulation of short-term U.S. Treasuries on associated yields, and aligning with financial models linking interest rates to shifts in demand for high-quality, short-term assets (Wolcott, 2020). The baseline models without controls show a weak positive semi-elasticity between Tether's market share and Treasury yields, with coefficients of 1.025 (0.887) for 1-month (3-month) yields in model 1 (4). However, adding a time trend reverses this relationship, resulting in negative and statistically significant semi-elasticities of -4.144 (-3.3765) for 1-month (3-month) yields in model 2 (5), demonstrating the importance of accounting for long-term macroeconomic and monetary trends to estimate the impact of Tether's market share.

The fully specified models, which control for both time trends and residualized issuance, report statistically significant semi-elasticities of -3.795 (-3.386) for 1-month (3-month) yields in models 3 (6). The estimates imply that a 1% increase in Tether's market share is associated with a yield reduction of 3.8% (3.4%), respectively. Given average yields of 4.16% in Q1 2025, this corresponds to a 14-16 basis point reduction in 1-month yields, reflecting the substantial influence of Tether's market share on short-term rates, depending on duration.[4][5] This economically significant effect aligns with traditional asset pricing theory as the direct liquidity impact of Tether's rapid U.S. Treasury bill accumulation—raising aggregate demand and lowering yields—and may further reflect broader market dynamics in line with the style investing theory of Barberis and Shleifer (2003), whereby large, predictable investors like Tether can amplify effects by creating style-level demand shifts as other market participants adjust portfolios based on the perceived stability and scale of Tether's recurring purchases.

## 3.2 Threshold Effects

To assess whether the relationship between Tether's market share and Treasury bill yields is nonlinear, we apply a threshold regression model based on Hansen (2000). This method estimates the point at which the effect of market share on yields shifts, capturing potential structural breaks. Using a grid search, we identify 0.973% as the optimal threshold, indicating a significant change in the relationship at this level.6 We thereby define the threshold regression model as:

$$\ln(y_t) = \alpha + \beta_1[(S_t) \times I(S_t \leq \tau)] + \beta_2[(S_t) \times I(S_t > \tau)] + \gamma T_t + \delta R_t + \varepsilon_t \qquad (2)$$

where $I(\cdot)$ is an indicator function equal to 1 if the condition is satisfied and 0 otherwise, while $\tau$ denotes the threshold value of 0.973%. The coefficients $\beta_1$ and $\beta_2$ measure the semi-elasticity of Treasury bill yields with respect to Tether's market share, capturing the percentage change in yields for a 1% change in market share. This specification permits the semi-elasticities of yields with respect to Tether's market share to vary across a low regime ($\leq$ 0.973%) or a high regime ($>$ 0.973%).

---

[4] The time trend coefficient remains positive and statistically significant across all specifications, capturing the broader upward movement in yields over the sample period, consistent with the Federal Reserve's aggressive rate-hiking cycle.

[5] Average yields are calculated from the raw yield data, not from the log-transformed values.

Table 3 presents the threshold regression results for 1-month and 3-month U.S. Treasury yields, using the estimated market share threshold of 0.973%. Figure 1 visualizes the corresponding fitted values for 1-month yields, highlighting the nonlinear relationship whereby Tether's impact on yields intensifies sharply once its market share crosses this threshold.7 For 1-month yields, the coefficient in the low regime (market share $\leq$ 0.973%) is -1.730, while in the high regime (market share $>$ 0.973%) it is -6.264. Tether's market share has a modest impact on yields when it remains below the threshold, but exerts stronger downward pressure as it passes this critical level. We yield analogous coefficients for 3-month yields of -1.419 in the low regime and -5.824 in the high regime.

To illustrate, in the low regime where Tether's market share is below or equal to 0.973% (e.g., a 0.1% increase in market share from 0.5% to 0.6%, equivalent to around $6.16 billion in 1-month Treasury bill purchases), reduces the log yield by 0.1730, or approximately 0.73 basis points at an average 4.24% yield level. In the high regime where Tether's market share is above 0.973%, the impact is larger in magnitude. A 0.1% increase in market share (e.g., 1% to 1.1%) reduces the log yield by 0.6264, roughly 2.6 basis points at the same yield level of 4.24%. These results reflect the nonlinear scaling effect: Tether's marginal influence on yields becomes much larger as its share of the market grows. For 3-month yields, the magnitudes are slightly smaller but comparable. In the low (high) regime, the coefficient registers -0.1419 (-0.5824), hence a 0.1% increase in Tether's market share reduces the log yield by around 0.59 (2.3) basis points at an average yield of 4.20%.

---

[6] The results imply that when Tether's market share reaches 0.973%, the effect on Treasury bill yields changes significantly. We assess this using a bootstrap likelihood ratio (LR) test, which produces an LR statistic of 34.59 and a bootstrap p-value of 0.0060 (based on 500 replications). This low p-value (p < 0.01) strongly rejects the null hypothesis of a purely linear relationship, providing strong evidence for the presence of a statistically significant threshold.

[7] The 3-month yield results exhibit a similar threshold pattern, with nearly identical visual characteristics, so we present only the 1-month yield figure for clarity.

The threshold behavior observed at 0.973% market share is economically intuitive and consistent with theoretical models of liquidity and price impact. At lower levels of market participation, Tether's U.S. Treasury bill purchases are likely absorbed by the deep liquidity of the sovereign debt market, resulting in modest statistically insignificant impact on yields. However, once Tether's holdings cross this critical level, further marginal purchases encounter an increasingly inelastic supply, as the market's ability to meet demand without affected price is reduced. This dramatic shift highlights that Tether-related demand pressure does not scale linearly but intensifies disproportionately as holdings grow, consistent with liquidity saturation and nonlinear price impact models (Kyle, 1985). As a result, downward yield pressure disproportionately amplifies as the available supply tightens, reflecting nonlinear scaling captured in the threshold estimates. Furthermore, this threshold behavior supports the theory of nonlinear market impact, particularly models emphasizing liquidity thresholds and order-flow saturation, as price effects intensify once critical volume thresholds are crossed (Gabaix et al., 2006; Kyle, 1985).

## 4  Conclusion

This paper presents the first empirical analysis of Tether's U.S. Treasury holdings on overall yields, offering a novel contribution to understanding how stablecoin reserves from the crypto sector influence traditional sovereign debt markets. Using a baseline semi-log trend model, we estimate Tether's market share of U.S. Treasury bills reduced 1-month yields by approximately 14-16 basis points relative to a counterfactual, resulting in roughly $10 billion in annual interest savings for the U.S. government.[8] Effects are similar for 3-month yields, with a semi-elasticity of 3.4%.

To account for potential nonlinearities, we apply a threshold regression model Hansen (2000) and identify a critical threshold at 0.973% market share using a grid search. Below this threshold—low regime—, Tether's influence on 1-month (3-month) yields is modest, with a 0.1% increase in market share reducing yields by about 0.73 (0.59) basis points, respectively. However, above this threshold (high regime), the same increase reduces yields by 2.6 (2.3) basis points, respectively, translating to roughly $14.8 billion ($13 billion) in annual interest savings. These results reflect the market's reduced capacity to absorb additional demand without notable price effects and highlight the nonlinear scaling of Tether's influence, where its marginal impact on yields grows disproportionately as its market share expands.

From a fiscal standpoint, this paper demonstrates that stablecoin demand can have beneficial budgetary implications as marginal increases in Tether's market share can reduce U.S. government financing costs by lowering sovereign yields. For investors, tracking stablecoin issuer reserves of U.S Treasuries can offer early signals of short-term yield movements, especially during periods of market volatility.

---

[8] Based on $6.2 trillion in outstanding U.S. Treasury bills with an average yield of 4.24%.

Furthermore, for policymakers, U.S. Treasury demand from stablecoin issuers can potentially absorb shocks from traditional finance, given the more heterogeneous stablecoin ownership comprising millions of entities and retail users, contrasting with more concentrated conventional U.S. Treasury investors. Sovereign debt market dependence on stablecoin issuer U.S. Treasury demand may complicate debt management and monetary policy, as holdings could fluctuate with regulatory changes or crypto-specific shocks unrelated to macroeconomic fundamentals. This aligns with concerns regarding balance sheet expansions by nonbank financial intermediaries (FSB, 2022), the stability of short-term funding markets (D'Avernas and Vandeweyer, 2024; Krishnamurthy and Li, 2023), and systemic risk implications of stablecoins Gorton and Zhang (2021). However, the retail-focused nature of stablecoin ownership may also buffer traditional financial markets against some shocks, potentially mitigating systemic risks.

While our analysis is limited to currently available data, it lays a foundation for future research as more detailed stablecoin issuer reserve information becomes available over time. As data on broader stablecoin reserve allocations emerge, future researchers can extend this work beyond U.S. Treasury bills to assess the full spectrum of asset impact, including indirect effects through general equilibrium mechanisms. Future studies should also explore longer time horizons to capture time-varying dynamics across different monetary regimes, providing a more comprehensive understanding of how stablecoin reserves influence traditional financial markets. This includes examining the broader effects on other fixed-income assets, potential spillovers between crypto markets and bond yields, and the ability of stablecoins to buffer shocks from traditional finance.

**Table 1: Summary Statistics and Correlations**

| Variable | Mean | SD | Min | Max | (i) | (ii) | (iii) | (iv) | (v) |
|---|---|---|---|---|---|---|---|---|---|
| **(i) Log 1-Month Yields** | 1.29 | 0.68 | -0.65 | 1.71 | | | | | |
| **(ii) Log 3-Month Yields** | 1.33 | 0.62 | -0.46 | 1.70 | 0.99 | | | | |
| **(iii) Tether's Market Share (%)** | 0.01 | 0.00 | 0.01 | 0.02 | 0.48 | 0.46 | | | |
| **(iv) Time Trend** | 6.50 | 3.45 | 1.00 | 13.00 | 0.65 | 0.62 | 0.64 | | |
| **(v) T-Bill Changes IHS** | 11.41 | 0.01 | 11.40 | 11.42 | 0.65 | 0.62 | 0.94 | 1.00 | |
| **(vi) T-Bill Changes Residual** | -0.00 | 0.04 | -0.07 | 0.05 | 0.30 | 0.32 | -0.07 | 0.00 | 0.01 |

Notes: Table 1 reports descriptive statistics for key variables from Q1 2022 to Q1 2025. (i) Log 1-Month Yields and (ii) Log 3-Month yields are natural logarithms of secondary market rates for newly issued 4-week and 13-week U.S. Treasury bills, sourced from the Federal Reserve Economic Data (FRED). (iii) Tether's Market Share (%) is the ratio of Tether's U.S. Treasury bill holdings, based on Consolidated Reserves Reports (CRRs), to total outstanding U.S. Treasury bills from U.S. Department of the Treasury data. (iv) Time Trend is a sequential variable from 1 (Q1 2022) to 13 (Q1 2025), capturing broader monetary conditions. (v) T-Bill Changes IHS measures monthly changes in outstanding U.S. Treasury bills, transformed using the inverse hyperbolic sine (IHS) function, $IHS(x) = \ln(x + \sqrt{x^2 + 1})$, to address large positive and negative values. (vi) T-Bill Changes Residual isolates the component of U.S. Treasury bill changes not linearly explained by the (iv) Time Trend or (iii) Tether's market share, calculated as the residuals from a regression of the (v) IHS-transformed U.S. Treasury bill changes on these variables, scaled by 1,000 for interpretability.

**Table 2: Baseline Time Trend Model Results**

|  | (1) Coef. (SE) | (2) Coef. (SE) | (3) Coef. (SE) | (4) Coef. (SE) | (5) Coef. (SE) | (6) Coef. (SE) |
|---|---|---|---|---|---|---|
| **(iii) Tether's Market Share** | 1.025* (0.555) | -4.144** (1.351) | -3.795** (1.402) | 0.887 (0.523) | -3.765** (1.347) | -3.386** (1.383) |
| **(iv) Time Trend** |  | 0.442*** (0.111) | 0.414*** (0.115) |  | 0.397*** (0.111) | 0.367** (0.114) |
| **(vi) T-Bill Change Residual** |  |  | 3.106 (3.204) |  |  | 3.382 (3.161) |
| **Constant** | 0.181 (0.623) | -2.654*** (0.743) | 2.473 (0.769) | 0.373 (0.585) | 2.599*** (0.741) | 2.400** (0.759) |
| **R2** | 0.235 | 0.704 | 0.732 | 0.208 | 0.654 | 0.693 |
| **Adj. R2** | 0.166 | 0.644 | 0.642 | 0.136 | 0.584 | 0.590 |
| **Dep. variable** | (i) 1-Month Yield | (i) 1-Month Yield | (i) 1-Month Yield | (ii) 3-Month Yield | (ii) 3-Month Yield | (ii) 3-Month Yield |

Notes: Table 2 presents the results of the baseline time trend model defined in Equation (1). The dependent variable is the natural logarithm of (i) 1-month U.S. Treasury bill yields in Models (1) to (3) and (ii) 3-month U.S. Treasury bill yields in Models (4) to (6). Models (1) and (4) are baseline specifications without controls. Models (2) and (5) include the (iv) Time Trend, while Models (3) and (6) incorporate both the (iv) Time Trend and the (vi) residualized T-bill issuance variable. Coefficients are reported with standard errors in parentheses. *, **, *** indicate statistical significance at the 10%, 5% and 1% levels respectively.

**Table 3: Threshold Regression Results**

|  | (1) Coef. (SE) | (2) Coef. (SE) |
| --- | --- | --- |
| **Intercept Shift (High Regime)** | 3.781*** | 3.603*** |
|  | (0.959) | (0.984) |
| **(iii) Tether's Market Share (≤0.973%)** | -1.730 | -1.419 |
|  | (1.015) | (1.041) |
| **(iii) Tether's Market Share (≥0.973%)** | -6.264*** | -5.824*** |
|  | (1.080) | (1.107) |
| **(iv) Time Trend** | 0.566*** | 0.526*** |
|  | (0.092) | (0.095) |
| **(vi) T-Bill Change Residual** | -0.531 | -0.074 |
|  | (2.196) | (2.251) |
| **Constant** | 0.394 | 0.390 |
|  | (0.701) | (0.719) |
| **R2** | 0.920 | 0.901 |
| **Adj. R2** | 0.862 | 0.829 |
| **Dep. variable** | (i) 1-Month Yield | (ii) 3-Month Yield |

Notes: Table 3 presents results of the threshold regression model defined in Equation (2). The optimal threshold of 0.973% is determined separately for each model via grid search. The dependent variable is the natural logarithm of (1) 1-month U.S. Treasury bill yields in decimal form in the first column and (2) 3-month U.S. Treasury bill yields in decimal form in the second column. The model includes separate coefficients for (iii) Tether's market share below (≤ 0.973%) and above (> 0.973%) the estimated threshold, capturing the nonlinear scaling effect described in the main text. Both models incorporate both the (iv) Time Trend and the (iv) residualized T-bill issuance variable. Coefficients are reported with standard errors in parentheses. *, **, *** indicate statistical significance at the 10%, 5% and 1% levels respectively.

**Figure 1. Threshold Regression Results for 1-Month T-Bill Yield**

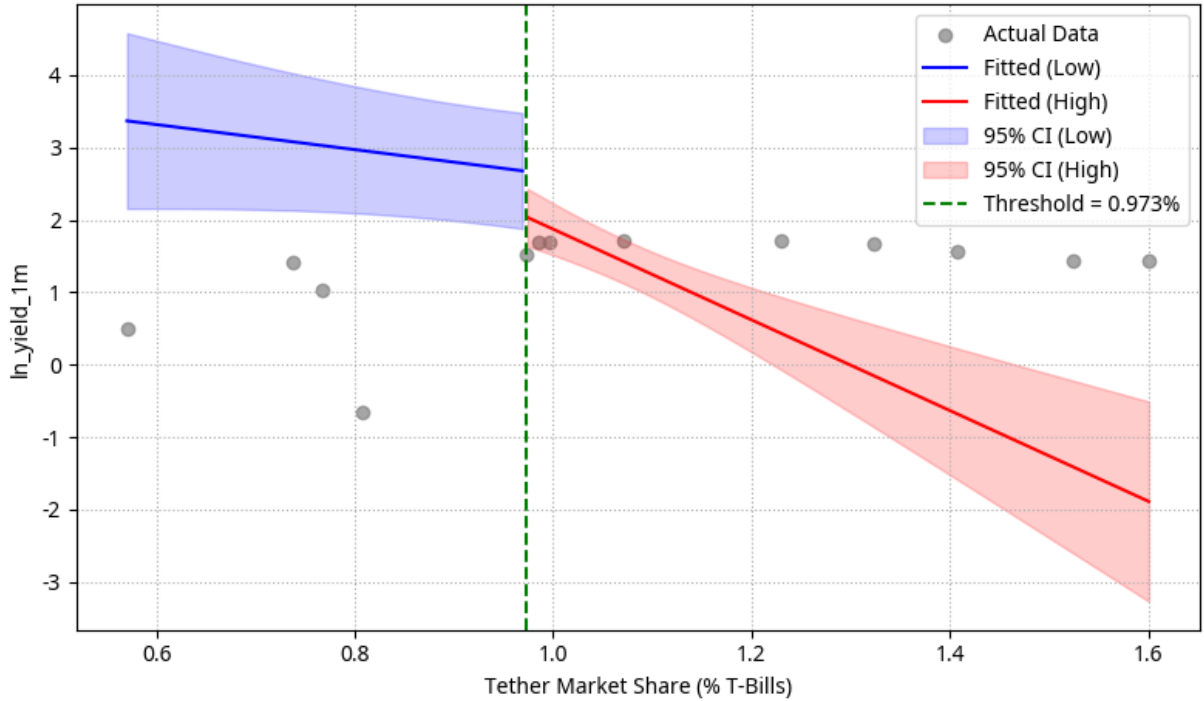

Notes: Figure 1 illustrates the results of the threshold regression model defined in Equation (2). The estimated optimal threshold of 0.973% is determined separately for each model via grid search, separating the sample into low and high regimes (indicated by the vertical green dashed line). The blue (low regime) and red (high regime) lines represent fitted regression lines and their respective 95% confidence intervals, utilizing the estimates of models (1) and (2) in Table 3. Gray dots represent actual data points, whose variation remains due to time trends and other controls not visualized. The dependent variable is the natural logarithm of (1) 1-month U.S. Treasury bill yields. The model includes separate coefficients for Tether's market share below ($\leq 0.973\%$) and above ($> 0.973\%$) the estimated threshold, capturing the nonlinear scaling effect described in the main text. The model incorporates both the (iii) Time Trend and the (iv) residualized T-bill issuance variable. Coefficients are reported with standard errors in parentheses. *, **, *** indicate statistical significance at the 10%, 5% and 1% levels respectively.